\documentclass[12pt]{article}

\usepackage{graphicx}
\usepackage{color}
\usepackage{epstopdf}
\usepackage{epsfig}
\usepackage{cite}
\usepackage{amssymb}
\usepackage{amsmath}
\usepackage{setspace}
\usepackage{float}

\begin{document}

\title{Nonsteady-state diffusion in two-dimensional periodic
  channels}

\author{Matan Sivan$^1$ and Oded Farago$^{1,2}$\\
$^1$Department of Biomedical Engineering, $^2$Ilse Katz Institute\\
for Nanoscale Science and Technology, Ben Gurion University,\\
Be'er Sheva 84105, Israel}

\maketitle
\begin{abstract}

  The dynamics of a freely diffusing particle in a two-dimensional
  channel with cross sectional area $A(x)$, can be effectively
  described by a one-dimensional diffusion equation under the action
  of a potential of mean force $U(x)=-k_BT\ln [A(x)]$ (where $k_BT$ is
  the thermal energy) in a system with a spatially-dependent diffusion
  coefficient $D(x)$. Several attempts to derive expressions relating
  $D(x)$ to $A(x)$ and its derivatives have been made, which were
  based on considering stationary flows in periodic channels. Here, we
  take an alternative approach and consider non-steady state single
  particle diffusion in an open periodic channel. The approach allows
  us to express $D(x)$ as a series of terms of increasing powers of
  $\epsilon$ - a parameter associated with the aspect ratio of the
  channel. When the expansion is truncated at the leading term, we
  recover the expression suggested by Zwanzig [J. Phys. Chem. {\bf
      96}, 3926 (1992)] for $D(x)$. Furthermore, comparison of the
  first few terms in our expansion for $D(x)$ with the one proposed by
  Kalinay and Percus [Phys. Rev. E {\bf 74}, 041203 (2006)] shows that
  they are consistent with each other. In the limit of long wavelength
  channels ($\epsilon\ll 1$), the expansion converges rapidly and the
  leading approximation provides a very accurate description of the
  two-dimensional dynamics. For short wavelength channels, the
  expansion does not converge and the validity of the effective
  one-dimensional description is questionable.
\end{abstract}

\newpage

\section{Introduction}

Transport of particles in narrow corrugated channels has received a
considerable attention in the past fifteen years. The reason for the
growing interest in problems of this type is their relevance to
numerous naturally occurring systems such as membrane ion
channel~\cite{hille}, carbon nanotubes~\cite{oconnell},
zeolites~\cite{schuring}, etc. Understanding molecular dynamics in
narrow channels is also important for many technological applications,
e.g., microfluidic devices~\cite{weigl} and solid state
nanopores~\cite{dekker}.

In this context, there have been numerous theoretical studies of the
following problem: Consider a particle moving in an open
two-dimensional channel whose long axis lies along the $x$-direction
($-\infty<x<\infty$). In the perpendicular $y$ direction, the channel
is bounded between $0\leq y\leq A(x)$, where $A(x)$ is a periodic
function with wavelength $\lambda$.  The two-dimensional (2D)
probability density, $\rho(x,y,t)$ (where $t$ denotes the time),
satisfies the diffusion equation
\begin{equation}
  \partial_t\rho(x,y,t)= D_0\nabla^2\rho(x,y,t)
  \label{eq:2diffusion}
\end{equation}
where $D_0$ is the medium diffusion
coefficient. Equation~(\ref{eq:2diffusion}) must be solved subject to
reflecting (Neumann) boundary conditions on the walls of the channel
\begin{equation}
  \partial_y\rho(x,y,t)|_{y=0}=0
  \label{eq:bc1}
\end{equation}
and
\begin{equation}
  \partial_y\rho(x,y,t)|_{y=A(x)}=A^{\prime}(x)\partial_x\rho(x,y,t).
  \label{eq:bc2}
\end{equation}

As the motion is limited to the longitudinal $x$-direction, one is
naturally interested in the one-dimensional (1D) probability density
function (PDF)
\begin{equation}
  P(x,t)=\int_0^{A(x)}\rho(x,y,t)\,dy.
  \label{eq:projection}
\end{equation}
It has been suggested that $P(x,t)$ may be found by solving the 1D
Smoluchowski equation describing Brownian dynamics under the action of
an entropic potential of mean force
$U(x)=-k_BT\ln\left[A\left(x\right)\right]$
\begin{equation}
  \frac{\partial P(x,t)}{\partial t}=D_0\frac{\partial}{\partial
    x}\left\{e^{-\beta U(x)}\frac{\partial}{\partial x}
  \left[\frac{P(x,t)}{e^{-\beta U(x)}}\right]\right\}
  =D_0\frac{\partial}{\partial
    x}\left\{A\left(x\right)\frac{\partial}{\partial x}
  \left[\frac{P(x,t)}{A\left(x\right)}\right]\right\},
  \label{eq:fickjacob}
\end{equation}
where $k_B$ is Boltzmann's constant, $T$ is the temperature, and
$\beta=(k_BT)^{-1}$. Eq.~(\ref{eq:fickjacob}) is known as Fick-Jacobs
(FJ) equation~\cite{jacobs}. Strictly speaking, the 1D description
provided by the FJ equations holds only when the two-dimensional
probability density is uniform along the $y$-direction, i.e., when
$\rho(x,y,t)=P(x,t)/A(x)$, which is generally not the case. Since the
agreement between the solution of FJ equation (\ref{eq:fickjacob}) and
simulation results may be quite poor~\cite{burada09,bdb15}, a modified
version of the FJ equation with a coordinate-dependent diffusion
coefficient, $D(x)$, has been considered
\begin{equation}
  \frac{\partial P(x,t)}{\partial t}=\frac{\partial}{\partial
    x}\left\{D(x)A(x)\frac{\partial}{\partial x}
  \left[\frac{P(x,t)}{A(x)}\right]\right\},
  \label{eq:fickjacob2}
\end{equation}
where $D(x)$ is supposedly a function of $A(x)$ and its derivatives
(see footnote~\cite{footnote3}).  Eq.~(\ref{eq:fickjacob2}) was first
derived by Zwanzig~\cite{zwanzig92}, by analyzing the temporal
evolution of the deviations in the local density from uniformity,
$\delta\rho(x,y,t)=\rho(x,y,t)-P(x,t)/A(x)$. From the analysis,
Zwanzig concluded that the introduction of a spatially-dependent
diffusion coefficient can improve the agreement between the 2D and 1D
descriptions, and he proposed the following expression for $D(x)$
\begin{equation}
D_{\rm Z}(x)=\frac{D_0}{1+A^{\prime\,2}(x)/3}.
\label{eq:zw}
\end{equation}
Notice that the reduction of the 2D diffusion equation
(\ref{eq:2diffusion}) to an effective 1D equation
(\ref{eq:fickjacob2}) cannot yield exact results since the 2D
diffusion process projected onto the $x$ direction is not Markovian
and thus cannot be described by a 1D diffusion equation with local
diffusion coefficient $D(x)$~\cite{mangeat17}.  Nevertheless,
Zwanzig's framework of Eq.~(\ref{eq:fickjacob2}) for depicting
transport in corrugated channels has become popular and other
expressions for $D(x)$ have been proposed, for instance,
\begin{equation}
  D_{\rm RR}(x)=\frac{D_0}{\left[1+A^{\prime\,2}(x)\right]^{1/3}},
  \label{eq:rr}
\end{equation}
and 
\begin{equation}
  D_{\rm
    KP}(x)=D_0\frac{\arctan\left[A^{\prime}(x)\right]}{A^{\prime}(x)},
  \label{eq:kp}
\end{equation}
which were suggested, respectively, by Reguera and Rubi
(RR)~\cite{reguera01} and by Kalinay and Percus
(KP)~\cite{kalinay06}. All the above three formulas for $D(x)$ (i)
satisfy $D(x)/D_0\sim 1-A^{\prime\,2}(x)/3$ for $A^{\prime}(x)\ll1$,
and (ii) ignore the higher order derivatives of $A(x)$. The latter
property of these expressions is not mathematically well justified. In
fact, the expression of KP (\ref{eq:kp}) was derived using a mapping
formalism~\cite{kalinay05} that generates a series of expressions that
include increasingly higher order derivatives of
$A(x)$. Unfortunately, the formalism involves a complicated
differential operator containing derivatives of all orders, which
makes it rather impractical. An alternative analytical approach has
been more recently introduced, which is based on formulation of the 2D
problem in the complex plane~\cite{kalinay14}. The derivation, via
this route, of expressions for $D(x)$ that take higher order
derivatives into account is still highly non-trivial; however, the
method has been recently exploited successfully for the derivation of
a series of expressions for the effective diffusion coefficient $D^*$
[see definition in Eq.~(\ref{eq:effectived}) below] in periodic
channels~\cite{mangeat17}.

In this paper, we consider dynamics in channels with periodic
cross-sectional area $A(x)$. We derive a series of approximations for
$D(x)$, successively taking into account higher order derivatives of
$A(x)$. In contrast to almost all previous studies of the problem (an
exception is ref.~\cite{bradley09}), our derivation is not based on
calculations of the steady-state PDF, but rather on the solution of
the time-dependent Smoluchowski equation with delta-function initial
conditions $P(x,0)=\delta(x)$. Similarly to~\cite{kalinay06}, each new
term in the series of expressions for $D(x)$ requires the calculation
of an exponentially increasing number of derivatives of functions of
$A(x)$; but in contrast to~\cite{kalinay06}, the differentiations that
need to be performed at each step are clearly expressed and not
formulated with differential operators that are hard to interpret. The
leading approximation coincides with Zwanzig's
formula~(\ref{eq:zw}). We explicitly give expressions for $D(x)$ up to
the fourth approximation involving the 8th derivative of $A(x)$.
Finally, we use computer simulations of a case study to evaluate the
importance of the higher order corrections. For slowly varying (long
wavelength) channels, the contribution of the higher order derivatives
appear to be rather small and unimportant, and the agreement between
the 1D and 2D simulations is excellent. As the periodicity decreases,
the higher order corrections may exhibit instability leading, locally,
to $D(x)>D_0$, and stronger deviations are found between the effective
diffusion coefficient computed in the 1D and 2D simulations.

The paper is organized as follows: In section \ref{sec:1dconst} we
summarize the main results of our recent work~\cite{sivan} for the PDF
of Brownian dynamics in a 1D periodic potentials, and in
section~\ref{sec:1dx} we extend these results to systems where the
friction coefficient is also periodic in space. In
Section~\ref{sec:2dchannels} we consider the 2D problem. First, in
section~\ref{sec:rho}, we derive an expression for the 2D density,
$\rho(x,y,t)$, in the form of an expansion in even powers in
$y$. Then, in section~\ref{sec:projected}, the 2D density is projected
onto the $x$ direction, and by comparison with the PDF derived in
section~\ref{sec:1dx}, we arrive at the expansion for $D(x)$ in
section~\ref{sec:diffconst}. The first few terms in this expansion are
calculated in section~\ref{sec:expansion}. In
section~\ref{sec:simulations}, we use computer simulations to test the
newly derived approximations for $D(x)$, and in
section~\ref{sec:summary} we summarize our results.

\section{Diffusion in a one-dimensional periodic potential}
\label{sec:1d}

\subsection{Constant diffusion coefficient}
\label{sec:1dconst}

We first consider FJ equation with constant $D_0$~(\ref{eq:fickjacob})
for a periodic channel with wavelength $\lambda$ and cross sectional
area $A(x)$. This equation represents an attempt to project the
two-dimensional diffusion equation onto the longitudinal $x$-axis by
introducing a periodic entropic potential of mean force
$U(x)=-k_BT\ln[A(x)/A_0]$, where $A_0=\langle A(x)
\rangle=(\lambda)^{-1}\int_0^{\lambda}A(x)dx$. In a previous
study~\cite{sivan}, we derived the general solution of this class of
diffusion equations subject to delta-function initial conditions,
$P(x,0)=\delta(x)$. We demonstrated that the PDF can be expressed as
an expansion of the following form
\begin{equation}
 P(x,t)=\frac{A(x)}{A_0}G(x,D^*t)\left[1+\sum_{n=1}^{\infty
   } q_n(x,t)\right]
 \label{eq:expansion}
\end{equation}
(see footnoe~\cite{footnote1}). In Eq.~(\ref{eq:expansion}),
$G(x,D^*t)=\exp(-x^2/4D^*t)/\sqrt{4\pi D^*t}$ is the normalized
Gaussian function, where $D^*\leq D_0$ is the effective diffusion
coefficient
\begin{equation}
  D^*=\lim_{t\to\infty}\frac{\left\langle x^2\right\rangle}{2t},
  \label{eq:effectived}
\end{equation}
which can be related to $U(x)$ via the Lifson-Jackson (LJ)
formula~\cite{lj}
\begin{equation}
  D^*=\frac{D_0}{\langle e^{-\beta U(x)}\rangle\langle e^{\beta U(x)}\rangle}=
  \frac{D_0}{A_0\langle[A(x)]^{-1}\rangle}.
  \label{eq:LJ}
\end{equation}
The terms $q_n(x,t)$ in Eq.~(\ref{eq:expansion}) are time-decaying
functions with asymptotic scaling behavior $q_n(x,t)\sim
t^{-n/2}$. These can be found by substituting the solution of the form
(\ref{eq:expansion}) into Eq.~(\ref{eq:fickjacob}), and comparing
terms with similar asymptotic scaling behavior on both sides. Notably,
we found that the leading time-decaying term, $q_1(x,t)=\lambda
xg(x)/4D^*t$ (see footnote~\cite{footnote2}), where $g(x)$ is a
periodic function given by
 \begin{equation}
   \lambda
   g(x)=2x-\frac{2}{\langle[A(x)]^{-1}\rangle}\,
   I\left[\frac{1}{A(x)}\right],
 \label{eq:g1}
 \end{equation}
and $I[f(x)]$ denotes the primitive function of $f(x)$ with $I(x=0)=0$.

\subsection{Coordinate-dependent diffusion coefficient}
\label{sec:1dx}

Let us now consider the same system, but with a space-dependent
diffusion periodic function, $D(x)$, with periodicity $\lambda$
similar to that of $A(x)$. We now need to solve the modified FJ
equation~(\ref{eq:fickjacob2}), the solution of which has the same
general form as in Eq.~(\ref{eq:expansion}). Focusing on the leading
time-decaying term, we write
\begin{equation}
 P(x,t)=\frac{A(x)}{A_0}G(x,D^*t)\left[1+ \frac{\lambda x g(x)}{4D^*t}\right],
 \label{eq:firstorder}
\end{equation}
which is correct up to order ${\cal O}(G/t^{1/2})\sim{\cal O}(1/t)$
provided that the correct function $g(x)$ is found. This is done by
substituting the solution~(\ref{eq:firstorder}) into
Eq.~(\ref{eq:fickjacob2}) and comparing terms that scale like
$G(x,D^*t)(x/t)$, which yields the following differential equation
\begin{equation}
 [D(x)A(x)(g^{\prime}-2)]^{\prime}=0.
 \label{eq:diffg1}
\end{equation}
Integrating this equation once with respect to $x$ gives
\begin{equation}
g^{\prime}=2+\frac{c_1}{A(x)D(x)}.
 \label{eq:diffg2}
\end{equation}
The constant $c_1$ can be determined by acknowledging that $g(x)$ is
periodic and, therefore,
 \begin{equation}
0=g(\lambda)-g(0)=\int_0^{\lambda} g^{\prime}(x) dx=2\lambda
+c_1\int_0^{\lambda} \frac{dx}{A(x)D(x)}.
 \label{eq:diffg3}
\end{equation}
Thus, the constant $c_1$ is given by:
\begin{equation}
  c_1=-\frac{2}{\langle[A(x)D(x)]^{-1}\rangle}.
\label{eq:c1}
\end{equation}
Integrating Eq.~(\ref{eq:diffg2}) once again with respect to $x$ gives
\begin{equation}
  g(x)=2x-\frac{2}{\langle[A(x)D(x)]^{-1}\rangle}
  I\left[\frac{1}{D(x)A(x)}\right],
  \label{eq:g2}
  \end{equation}
which reduces to (\ref{eq:g1}) when $D(x)=D_0$. Notice that when the
diffusion coefficient depends on the coordinate $x$, the effective
diffusion coefficient $D^*$ is given by the modified Lifson Jackson
(MLJ) formula~\cite{zwanzig92}
\begin{equation}
  D^*=\frac{1}{A_0\langle[D(x)A(x)]^{-1}\rangle},
  \label{eq:MLJ}
\end{equation}
which reduces to (\ref{eq:LJ}) when $D(x)=D_0$, and allows one to also
write
\begin{equation}
  g(x)=2x-2A_0 D^*
  I\left[\frac{1}{D(x)A(x)}\right],
  \label{eq:g3}
\end{equation}

\section{Diffusion in a two-dimensional periodic channel}
\label{sec:2dchannels}

\subsection{The two-dimensional density}
\label{sec:rho}

In section \ref{sec:1dx} we presented the solution of the modified FJ
equation for a {\it given}\/ diffusion function $D(x)$. It is given by
Eq.~(\ref{eq:firstorder}) [with $D^*$ and $g(x)$ given by
  Eqs.~(\ref{eq:MLJ}) and (\ref{eq:g3}), respectively], and it is
correct up to order $\sim{\cal O}(1/t)$ at large times. The goal now
is to find an expression for $D(x)$, for which this solution provides
the best approximation to the true projected PDF. The latter is
obtained, via Eq.~(\ref{eq:projection}), from the 2D density,
$\rho(x,y,t)$, that solves the 2D diffusion
equation~(\ref{eq:2diffusion}) with the reflecting boundary conditions
(\ref{eq:bc1}) and (\ref{eq:bc2}) at the walls of the channel. The
fact that the projected 1D PDF takes the asymptotic (large $t$) form
of Eq.~(\ref{eq:firstorder}) implies that the 2D density has following
asymptotic form
\begin{equation}
  \rho(x,y,t)=\frac{G(x,D^*t)}{A_o}
  \left[1+\frac{\lambda x}{4D^*t}\sum_{n=0}^{\infty}f_n(x)y^{2n}\right],
\label{eq:rho}
\end{equation}
where $f_n(x)$ are some functions to be determined. Only even powers
of $y$ are included in this expression due to the invariance of the
equation and the boundary conditions with respect to reflection around
the $x$ axis ($y\leftrightarrow -y$).

We note here that although Eqs.~(\ref{eq:firstorder}) and
(\ref{eq:rho}) represent solutions that are only asymptotically
correct and that they miss higher order time-decaying terms, these
forms are sufficient for the sake of the task in hand which is to find
the best choice of $D(x)$. This is because, as hinted by
Eq.~(\ref{eq:g2}), the information needed for the determination of
$D(x)$ is encompassed in the leading asymptotic correction.

We proceed by first noting that expression (\ref{eq:rho}) satisfies
automatically the boundary condition (\ref{eq:bc1}) at $y=0$. We then
follow a route similar to the one presented in section \ref{sec:1dx}
for the determination of $g(x)$, and substitute expression
(\ref{eq:rho}) in Eq.~(\ref{eq:2diffusion}).  By comparing terms of
the form $G(x,D^*t)(xy^{2n}/t)$ on both sides of the equation we
arrive at the following recurrence relation (for $n\geq 1$)
\begin{equation}
  f_{n-1}^{\prime\prime}(x)+2n(2n-1)f_n(x)=0,
  \label{eq:rec1}
\end{equation}
which can be successively solved to yield
\begin{equation}
  f_n(x)=\frac{(-1)^n}{(2n)!}f_0^{(2n)}(x),
  \label{eq:rec2}
\end{equation}
where (throughout this paper) $f_0^{(m)}(x)$ denotes the derivative of
order $m$ of the function $f_0(x)$.

The function $f_0$ [from which all other functions $f_n$ can be
  derived via relation (\ref{eq:rec2})] can be found from the
remaining boundary condition at $y=A(x)$. This is done by substituting
expression (\ref{eq:rho}) in (\ref{eq:bc2}) and, again, comparing only
terms proportional to $G(x,D^*t)(x/t)$. This leads to the following
equation
\begin{equation}
  \sum_{n=1}^{\infty}2nf_n(x)\left[A(x)\right]^{2n-1}
  =A^{\prime}(x)\left\{-2+\sum_{n=0}^{\infty}
  f_n^{\prime}(x)\left[A(x)\right]^{2n} \right\},
\label{eq:bc11}
\end{equation}
which by using relation (\ref{eq:rec2}) can be also written in the
form
\begin{equation}
  \sum_{n=1}^{\infty}\frac{(-1)^nf_0^{(2n)}(x)\left[A(x)\right]^{2n-1}}{(2n-1)!}
  =A^{\prime}(x)\left\{-2+\sum_{n=0}^{\infty}
  \frac{(-1)^nf_0^{(2n+1)}(x)\left[A(x)\right]^{2n}}{(2n)!} \right\},
\label{eq:bc12}
\end{equation}
involving only the function $f_0(x)$. Finally, we define the function
\begin{equation}
\psi(x)=f_0^{\prime}(x)-2,
\label{eq:psi}
\end{equation}
and rewrite Eq.~(\ref{eq:bc12}) as
\begin{equation}
  \sum_{n=1}^{\infty}\frac{(-1)^n\psi^{(2n-1)}(x)\left[A(x)\right]^{2n-1}}{(2n-1)!}
  =A^{\prime}(x)\left[\sum_{n=0}^{\infty}
    \frac{(-1)^n\psi^{(2n)}(x)\left[A(x)\right]^{2n}}{(2n)!} \right].
\label{eq:bc13}
\end{equation}

\subsection{The projected one-dimensional PDF}
\label{sec:projected}

By using Eqs.~(\ref{eq:rho}) and (\ref{eq:rec2}) in
Eq.~(\ref{eq:projection}) we arrive at
\begin{equation}
  P(x,t)=\int_0^{A(x)}\rho(x,y,t)dy=\frac{G(x,D^*t)A(x)}{A_0}
  \left\{1+\frac{\lambda
    x}{4D^*t}\sum_{n=0}^{\infty}\frac{(-1)^nf_0^{(2n)}(x)\left[A(x)\right]^{2n}}
                {(2n+1)!}  \right\}.
\label{eq:project1}
\end{equation}
Comparing Eq.~(\ref{eq:project1}) to Eq.~(\ref{eq:firstorder}) leads
to
\begin{equation}
g(x)=\sum_{n=0}^{\infty}\frac{(-1)^nf_0^{(2n)}(x)\left[A(x)\right]^{2n}}{(2n+1)!}.
\label{eq:gproject}
\end{equation}

\subsection{The spatially-dependent diffusion coefficient}
\label{sec:diffconst}

The coordinate-dependent diffusivity $D(x)$ can be now identified by
writing the function $g(x)$ in Eq.~(\ref{eq:gproject}) in the form of
Eq.~(\ref{eq:g3}). Thus, we wish to find the function $D(x)$
satisfying
\begin{equation}
  \sum_{n=0}^{\infty}\frac{(-1)^nf_0^{(2n)}(x)\left[A(x)\right]^{2n}}{(2n+1)!}
  =2x-2A_0D^* I\left[\frac{1}{D(x)A(x)}\right].
\label{eq:gproject2}
\end{equation}
Differentiating Eq.~(\ref{eq:gproject2}) with respect to $x$ gives
\begin{equation}
  \frac{2A_0D^*}{A(x)D(x)}=2-\left\{\sum_{n=0}^{\infty}
  \frac{(-1)^nf_0^{(2n)}(x)\left[A(x)\right]^{2n}}{(2n+1)!}\right\}^{\prime},
\label{eq:gproject3}
\end{equation} 
and by using Eq.~(\ref{eq:psi}) we may also write
\begin{equation}
  \frac{2A_0D^*}{A(x)D(x)}=-\psi(x)-\left\{\sum_{n=1}^{\infty}
  \frac{(-1)^n\psi^{(2n-1)}(x)\left[A(x)\right]^{2n}}{(2n+1)!}\right\}^{\prime}.
\label{eq:gproject4}
\end{equation} 
By reciprocating Eq.~(\ref{eq:gproject4}) we finally arrive at
\begin{equation}
  D(x)=-\frac{2A_0D^*}{A(x)\left[\psi(x) +\left\{\sum_{n=1}^{\infty}
      \frac{(-1)^n\psi^{(2n-1)}(x)\left[A(x)\right]^{2n}}
           {(2n+1)!}\right\}^{\prime}\right]}.
\label{eq:gproject5}
\end{equation}

In order to find $D(x)$ from Eq.~(\ref{eq:gproject5}), we now need to
find the function $\psi(x)$ by solving
Eq.~(\ref{eq:bc13}). Unfortunately, this equation involves
derivatives of $\psi(x)$ of any order and, therefore, cannot be
solved. In what follows we present a set of approximations for
$\psi(x)$ and $D(x)$. Notice, that Eq.~(\ref{eq:bc13}) is a
homogeneous differential equation and, therefore, the function
$\psi(x)$ can be determined up to a multiplicative
constant. Therefore, we will rewrite Eq.~(\ref{eq:gproject5})
\begin{equation}
  D(x)=\frac{A_0{\cal D}}{A(x)\left[\psi(x) +\left\{\sum_{n=1}^{\infty}
      \frac{(-1)^n\psi^{(2n-1)}(x)\left[A(x)\right]^{2n}}
           {(2n+1)!}\right\}^{\prime}\right]},
\label{eq:gproject6}
\end{equation}
where ${\cal D}$ is some diffusion coefficient that depends on the
choice of the multiplicative constant in the definition of
$\psi(x)$. The diffusion constant ${\cal D}$ will be determined by
other considerations.

\subsection{Series expansion}
\label{sec:expansion}

The function $\psi(x)$ is periodic with wavelength
$\lambda$. Introducing the dimensionless parameter
$\epsilon=A_0/\lambda\sim A^{\prime}$ which becomes vanishingly small
for narrow and slowly varying channels, we can formally write the
function $\psi(x)$ as an expansion in terms of increasing orders of
$\epsilon$, namely
\begin{equation}
  \psi(x)=\psi_0(x)+\psi_1(x)+\psi_2(x)+\cdots\,\, ,
  \label{eq:psiexpansion}
\end{equation}
where
\begin{equation}
\psi_n(x)\sim \epsilon^{2n}
\label{eq:orderk}
\end{equation}
The scaling behavior (\ref{eq:orderk}) follows from the fact to be
shown henceforth that $\psi_{n+1}\sim A^2\psi_n^{\prime\prime}\sim
A^2\psi_n/\lambda^2\sim \epsilon^2\psi_n$.

In order to obtain the $n$-th function $\psi_n(x)$, we need to
identify the terms in Eq.~(\ref{eq:bc13}) of order $\epsilon^{2n+1}$.
 
{\bf The Zeroth approximation.} In this approximation
$\psi(x)=\psi_0(x)$, and only the first terms in the sums on both
sides of Eq.~(\ref{eq:bc13}) are kept. Thus we have the equation
\begin{equation}
  -\psi_0^{\prime}(x)A(x)=\psi_0(x)A^{\prime}(x),
  \label{eq:order0}
\end{equation}
with terms of order $\sim \epsilon$ on both sides, and with the
solution
\begin{equation}
  \psi_0(x)=\frac{A_0}{A(x)}.
\label{eq:order01}
\end{equation}
Notice that in order to keep the function $\psi$ dimensionless, we
pick $A_0$ as our choice for the ``arbitrary'' multiplicative constant
in its definition [see discussion around Eq.~(\ref{eq:gproject6})
  above].

The zeroth approximation of $D(x)$ is obtained by substituting $\psi=
\psi_0(x)$ in Eq.~(\ref{eq:gproject6}) and keeping only the leading
term in the square brackets in the denominator. This gives
\begin{equation}
  D(x) = \frac{A_0{\cal D}}{A(x)\psi_0(x)}={\cal D}.
  \label{eq:order02}
\end{equation}
Since the zeroth approximation of $D(x)$ must converge to the correct
value in the limit $\epsilon\rightarrow 0$, which corresponds to the
case of a flat channel, we must set ${\cal D}=D_0$. Thus, to zero
order in $\epsilon$
\begin{equation}
  D(x) = D_0,
  \label{eq:order03}
\end{equation}
which upon substitution in the modified FJ equation
(\ref{eq:fickjacob2}), reduce it to the form of the original FJ
equation~(\ref{eq:fickjacob}).

{\bf The first correction:} To a first approximation,
$\psi(x)=\psi_0(x)+\psi_1(x)$, where $\psi_0(x)$ is given by
Eq.~(\ref{eq:order01}).  The function $\psi_1(x)$ is found by solving
the following equation
\begin{equation}
-\psi_1^{\prime }(x)A(x)+\frac
{1}{3!}\psi_0^{(3)}(x)A^3(x)=A^{\prime}(x)\left[\psi_1(x)
  -\frac{1}{2!}\psi_0^{\prime
    \prime}(x)A^2(x)\right].
\label{eq:order1}
\end{equation}
This equation is derived by: (i) writing Eq.~(\ref{eq:bc13}) for
$\psi=\psi_0+\psi_1$ with only two terms in each sum on both sides
(i.e., one term more than in the zeroth approximation), and (ii)
isolating the terms that scale $\sim\epsilon^3$.  (The terms scaling
$\sim \epsilon$ constitute the already solved Eq.~(\ref{eq:order0}),
and the terms scaling $\sim\epsilon^5$ are discarded.)  Since
Eq.~(\ref{eq:order1}) can be also written as
\begin{equation}
  \left[\psi_1(x)A(x)\right]^{\prime}=
  \frac{\left[\psi_0^{\prime \prime }(x)A^3(x)\right]^{\prime}}{3!},
\end{equation}
we immediately find that
\begin{equation}
\psi_1(x)=\frac{\psi_0^{\prime \prime }(x)A^2(x)}{3!}.
\label{eq:order11}
\end{equation}

Let us denote the $n$-th approximation of $D(x)$ by $D_n(x)$. We
already found that $D_0(x)=D_0$ [see Eq.(\ref{eq:order03})]. The first
approximation, $D_1(x)$, is derived by substituting
$\psi(x)=\psi_0(x)+\psi_1(x)$ in the leading term in the square
brackets, and $\psi(x)=\psi_0(x)$ in the first term in the sum
($n=1$). Thus, to first approximation
\begin{equation}
D_1(x)=\frac{A_0D_0}{A(x)\left[\left\{\psi_0(x)+\psi_1(x)\right\}-
\left\{\psi_0^{\prime}(x)A^2(x)/3!\right\}^{\prime}\right]},
\label{eq:order12}
\end{equation}
which is correct to order $\epsilon^2$. (In general, $D_n(x)$, is
correct to order $\epsilon^{2n}$.) By using Eq.~(\ref{eq:order11}), we
can also write the alternative form for (\ref{eq:order12})
\begin{equation}
  D_1(x)=\frac{A_0D_0}{A(x)\left[\psi_0(x)
      -2\psi_0^{\prime}(x)A(x)A^{\prime}(x)/3!\right]}.
\label{eq:order13}
\end{equation}
By using Eq.~(\ref{eq:order01}) in Eq.~({\ref{eq:order13}) we arrive
  at
\begin{equation}
D_1(x)=\frac{D_0}{1+\left[A^{\prime}(x)\right]^2/3},
\label{eq:order14}
\end{equation}
which is the Zwanzig formula (\ref{eq:zw}). 

{\bf The second approximation:} Similarly, the second approximation
for the function $\psi$ reads $\psi=\psi_0+\psi_1+\psi_2$, and the
latter term can be found from the equation for the terms in
(\ref{eq:bc13}) scaling $\sim\epsilon^5$. The equation reads
\begin{equation}
-\psi_2^{\prime }A+\frac
{1}{3!}\psi_1^{(3)}A^3-\frac{1}{5!}\psi_0^{(5)}A^5=
A^{\prime}\left[\psi_2-\frac{1}{2}\psi_1^{\prime
    \prime}A^2+\frac{1}{4!}\psi_0^{(4)}A^4\right],
\label{eq:order2}
\end{equation}
which can be also written as
\begin{equation}
\left( \psi_2A\right)^{\prime}=\left[-\frac{1}{5!}\psi_0^{(4)}A^5+
  \frac{1}{3!}\psi_1^{\prime \prime }A^3\right]^{\prime}.
\label{eq:order21}
\end{equation}
Thus,
\begin{equation}
\psi_2=-\frac{1}{5!}\psi_0^{(4)}A^4+ \frac{1}{3!}\psi_1^{\prime \prime }A^2.
\label{eq:order22}
\end{equation}
The second approximation, $D_2(x)$, is derived by truncating the sum
in the denominator at $n=2$, and keeping only terms up to order
$\epsilon^4$. This yields,
\begin{equation}
  D_2(x)=\frac{A_0D_0}{A\left[\left\{\psi_0 +\psi_1+\psi_2\right\}-
      \left\{\left(\psi_0^{\prime}+\psi_1^{\prime}\right)
      A^2/3!\right\}^{\prime}+\left\{\psi_0^{(3)}A^4/5!\right\}^{\prime}\right]}.
\label{eq:order23}
\end{equation}
Using Eqs.~(\ref{eq:order01}), (\ref{eq:order11}), and
(\ref{eq:order22}) in Eq.~(\ref{eq:order23}), we arrive at the second
approximation for $D(x)$ 
\begin{equation}
D_2(x)= \frac{D_0}{1+A^{\prime \,
    2}/3+(A^2A^{\prime}A^{(3)}-AA^{\prime \, 2}A^{\prime
    \prime}-4A^{\prime \, 4})/45}
\label{eq:order24}
\end{equation}

{\bf Higher order corrections} Following the same scheme, one can
readily find that the $\psi_n$ can be obtained recursively via the
relation
\begin{equation}
\psi_n=\sum_{k=0}^{n-1}\frac{(-1)^{n-k-1}\psi_k^{(2n-2k)}A^{2n-2k}}{(2n-2k+1)!},
\label{eq:psi_n}
\end{equation}
with $\psi_0=A_0/A$ (\ref{eq:order01}). The $n$-th approximation,
$D_n(x)$ is given by
\begin{equation}
  D_n(x)=\frac{A_0D_0}{A\left[\sum_{k=0}^{n}\psi_k+\sum_{k=1}^{n}
      \frac{(-1)^k}{(2k+1)!}
      \left\{\sum_{l=0}^{n-k}\psi_l^{(2k-1)}A^{2k}\right\}^{\prime}\right]}.
    \label{eq:ordern}
\end{equation}
A more ``user-friendly'' expression can be wrtiten for the $n$-th
approximation of the friction coefficient $1/D_n(x)$
\begin{equation}
  \frac{1}{D_n(x)}=\frac{A}{A_0D_0}\left[\sum_{k=0}^{n}\psi_k+\sum_{k=1}^{n}
    \frac{(-1)^k}{(2k+1)!}
    \left\{\sum_{l=0}^{n-k}\psi_l^{(2k-1)}A^{2k}\right\}^{\prime}\,\right],
    \label{eq:ordernfric}
\end{equation}
which can be decomposed into two contributions as follows:
\begin{eqnarray}
\frac{1}{D_n(x)}&=&\frac{A}{A_0D_0}
\left[\sum_{k=0}^{n-1}\psi_k+\sum_{k=1}^{n-1} \frac{(-1)^k}{(2k+1)!}
  \left\{\sum_{l=0}^{(n-1)-k}\psi_l^{(2k-1)}A^{2k}\right\}^{\prime}\,\right]
      \label{eq:ordern1}\\
      &+&\frac{A}{A_0D_0}\left[\psi_n+
        \left\{\sum_{k=1}^{n}\frac{(-1)^k}{(2k+1)!}
        \psi_{n-k}^{(2k-1)}A^{2k}\right\}^{\prime}\,\right]\nonumber\\
      &=&\frac{1}{D_{n-1}(x)}+\frac{A}{A_0D_0}\left[\psi_{n}+
        \left\{\sum_{k=1}^{n}\frac{(-1)^k}{(2k+1)!}
        \psi_{n-k}^{(2k-1)}A^{2k}\right\}^{\prime}\,\right].\nonumber
\end{eqnarray}
By using Eq.~({\ref{eq:psi_n}) and changing the index of summation in
  (\ref{eq:ordern1}) from $k$ to $l=n-k$, we arrive at
\begin{equation}
  \frac{1}{D_n}=\frac{1}{D_{n-1}}+\frac{AA^{\prime}}{A_0D_0}
  \sum_{l=0}^{n-1}\frac{(-1)^{n-l}(2n-2l)}{(2n-2l+1)!}\psi_l^{(2n-2l-1)}A^{2n-2l-1}.
\label{eq:ordern_recursive}
\end{equation}
Using Eqs.~(\ref{eq:psi_n}) and (\ref{eq:ordern_recursive}), we
calculate the third approximation
\begin{eqnarray}
  \frac{1}{D_3}&=&\frac{1}{D_2}+\frac{1}{945D_0}
  \left[2A^4A^{\prime}A^{(5)}+8A^3A^{\prime \, 2}A^{(4)}
    -12A^3A^{\prime}A^{\prime \prime}A^{(3)}\right. 
 \label{eq:D3} \\
&-& \left. 27A^2A^{\prime \, 3}A^{(3)} -58A^2A^{\prime \, 2}A^{\prime
   \prime \,2} \nonumber +31AA^{\prime \, 4}A^{\prime
   \prime}+44A^{\prime \, 6}\right],
\end{eqnarray}
and the fourth approximation
\begin{eqnarray}
  \frac{1}{D_4}&=&\frac{1}{D_3}+\frac{1}{14175D_0}
  \left[3A^6A^{\prime}A^{(7)}+39A^5A^{\prime
    \, 2}A^{(6)}+5A^5A^{\prime}A^{\prime \prime}A^{(5)}\right. 
 \label{eq:D4}\\
&+& 74A^4A^{\prime \, 3}A^{(5)}-53A^5A^{\prime }A^{(3)}A^{(4)}
 -412A^4A^{\prime \,2}A^{\prime
   \prime}A^{(4)}-118A^3A^{\prime \, 4}A^{(4)} \nonumber \\
 &-& 911A^4A^{\prime
   \, 2 }[A^{(3)}]^2-682 A^3A^{\prime \,3 }A^{\prime \prime
 }A^{(3)}+451 A^2A^{\prime
   \,5}A^{(3)}-467A^4A^{\prime}A^{\prime \prime \,2} A^{(3)} \nonumber \\
  &+& \left. 157A^3A^{\prime \,2}A^{\prime\prime \,3
 }+1956A^2A^{\prime \,4}A^{\prime \prime \, 2} -555AA^{\prime \,6}
 A^{\prime \prime}-428A^{\prime \,8} \right] \nonumber.
\end{eqnarray}

In principle one can proceed and derive the higher order
approximations $D_n(x)$ in the same manner, but in practice the number
of differentiations that need to be carried grows exponentially with
$n$ and the calculations become tedious. The same feature complicates
the calculation of the series of $D_n(x)$ in ref.~\cite{kalinay06},
but the approach in that work ``suffers'' from an extra complication
which is the use of differential operators containing derivatives of
all orders that are very hard to identify. The use of
Eqs.~(\ref{eq:psi_n}) and (\ref{eq:ordern_recursive}) clearly offers a
far more tractable route to finding the higher order terms. The
expressions for $D_1$, $D_2$, and $D_3$ given here by
Eqs.~(\ref{eq:order14}), (\ref{eq:order24}), and (\ref{eq:D3}),
respectively, are different from their counterparts
in~ref.~\cite{kalinay06} [see Eq.~(13) therein]. However, if we Taylor
expand the former and leave in the expansion only terms up to order
$\epsilon^{2n}$ than the latter are recovered. It is reasonable to
speculate that this also holds true for $n>3$.

\section{Simulation results}
\label{sec:simulations}

As a case study, we consider diffusion in a channel with cross
sectional area given by $A(x)=h_0+\Delta [(2x/\lambda)^2-1]^2$ for
$x\in[-\lambda/2,\lambda/2]$ and repeated periodically outside of this
interval. We set the parameters $h_0=3$ (minimum channel opening) and
$\Delta=12$ (amplitude of channel height oscillations), and take the
channel periodicity to be either $\lambda=90$ or $\lambda=30$. The
average height of the channel is $A_0=\langle
A(x)\rangle=8\Delta/15+h_0= 9.4$. Therefore, the corresponding values
of $\epsilon= A_0/\lambda$ are 0.10 and 0.31 for $\lambda=90$ and
$\lambda=30$, respectively. We set the medium diffusion coefficient
$D_0$ to unity. Fig.~\ref{fig:fig1} shows the first three
approximations $D_n(x)$ [$n=1,2,3$ in Eqs.~(\ref{eq:order14}),
  (\ref{eq:order24}), and (\ref{eq:D3}), respectively], as well as the
expression of Kalinay and Percus (KP) [Eq.~(\ref{eq:kp})] which is the
limit ($n\rightarrow\infty$) expression when all the derivatives of
$A(x)$, except for the first one, are set to zero. In
Fig.~\ref{fig:fig1}~(a) we plot the diffusion functions corresponding
to the channel with the long wavelength $\lambda=90$. All the
expressions look remarkably identical, which is not surprising since
the variable $\epsilon$ in the power expansion in section
\ref{sec:expansion} is indeed small in this case and the higher order
corrections are expected to vanish rapidly. In contrast, for the case
of a short wavelength channel with $\lambda=30$ depicted in
Fig.~\ref{fig:fig1}~(b), significant variations between the different
expressions are observed. This is the regime where for some values of
$x$, $|A^{\prime}(x)|>1$,
and the series expansion fails to converge. Particular notice should
be given to the fact that truncating the expansion at a finite $n$ may
locally lead to $D(x)>D_0=1$ [see, e.g., $D_3(x)$ in
  Fig.~\ref{fig:fig1}~(b)], which best demonstrates that the higher
order corrections may become increasingly large.

\begin{figure}[t]
\centering\includegraphics[width=0.4\textwidth]{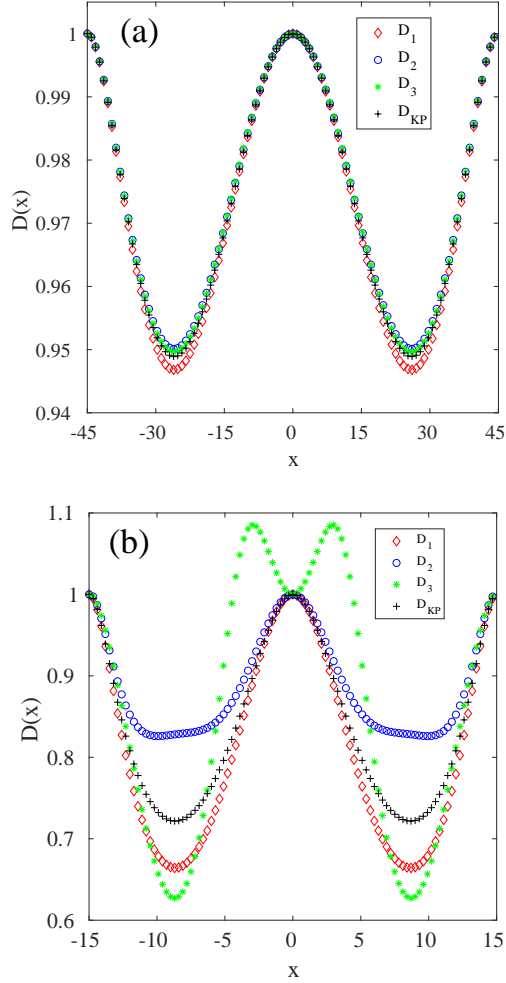} 
\caption{(Color Online) The coordinate-dependent diffusion
  coefficients $D_1$ (red diamonds), $D_2$(blue circles), $D_3$(green
  stars), and KP expression $D_{\rm KP}$ (black pluses) for the case
  studies discussed in the text. The periodicity of the channel is
  $\lambda=90$ ($\epsilon=0.10$) in (a) and $\lambda=30$
  ($\epsilon=0.31$) in (b)}.
\label{fig:fig1}
\end{figure}

To further test the accuracy of the 1D effective description of the
dynamics, we performed Langevin dynamics simulations of both 2D
channels with a height profile $A(x)$ and a constant diffusion
coefficient $D_0=1$ (case 1), and of 1D systems with a periodic
potential $U(x)=-k_BT\ln\left[A\left(x\right)\right]$ and various
periodic diffusion functions $D(x)$, including $D_0=1$ (case 2), our
expressions for $D_1$, $D_2$, $D_3$ (cases 3-5), and $D_{\rm KP}$
(case 6). In each case, we measured the effective diffusion
coefficient $D^*$ via Eq.~(\ref{eq:effectived}), by simulating
$5\times 10^8$ long trajectories of particles starting at the
origin. The trajectories were computed using the G-JF integrator for
Langevin's equation of motion~\cite{gjf}, and the spatial variations
in $D(x)$ were accounted for by setting the value of the friction
coefficient, $\alpha(x)=k_BT/D(x)$, corresponding to each time step
according to the recently proposed ``inertial
convention''~\cite{gjf2}. This combination (of an integrator and
convention for handling the multiplicative noise) produces excellent
results even for relatively large integration time steps. Our results
are summarized in tables~\ref{tab:lambda90} and \ref{tab:lambda30}
which give the simulation values, along with the corresponding values
derived from the MLJ formula (\ref{eq:MLJ}) for 1D periodic
systems. Table.~\ref{tab:lambda90} shows the results for a channel of
wavelength $\lambda=90$. As can be deduced from the table, the first
order diffusion coefficient $D_1(x)$ [Zwanzig's expression
  (\ref{eq:zw})] gives remarkably precise results when compared to the
2D case. The higher order corrections, which in
Fig.~\ref{fig:fig1}~(a) appear rather small, are unnecessary and do
not yield any further improvement in the results. Furthermore, the
data confirms that for slowly varying channels, The MLJ formula gives
values of $D^*$ that are identical to the numerical counterparts (see
discussion in~\cite{sivan}). In contrast, table~\ref{tab:lambda30}
reveals that for $\lambda=30$, the agreement of the 1D simulations
results with the 2D case is far from perfect. The first order
diffusion coefficient $D_1$ exhibits significant improvement compared
to the zeroth approximation $D_0$, but $D_2$ is no better than
$D_1$. Both $D_3$ and $D_{\rm KP}$ appear to give $D^*$ that is quite
close to the value measured in the 2D simulations, but this is clearly
coincidental. Once the series expansion fails to converge (as
demonstrated by the strong variations between the successive
approximations exhibited in Fig~\ref{fig:fig1}), the accuracy of the
1D effective picture of the dynamics becomes highly questionable. The
MLJ results follow the trends exhibited by their corresponding
numerical values, albeit with a lesser degree of accuracy than for
$\lambda=90$.

\begin{table}[t]
  \caption{Effective diffusion coefficients for channels with $\lambda=90$}
\centering
\begin{tabular}{l c c}
  \hline\hline
Case studied & Simulation results & MLJ formula \\ [0.5ex]
\hline
(1) 2-dim& 0.7340(5) & - \\
(2) $D_0$ &  0.7543(5) & 0.7564 \\
(3) $D_1 $& 0.7344(3) &0.7367  \\
(4) $D_2$ & 0.7353(5) &  0.7377\\
(5) $D_3$ &0.7351(3) &  0.7376\\ 
(6) $D_{\rm KP}$ &0.7346(3) & 0.7372\\[1ex]
\hline
\end{tabular}
\label{tab:lambda90}
\end{table}

\begin{table}[b]
\caption{Effective diffusion coefficients for channels with $\lambda=30$}
\centering
\begin{tabular}{l c c }
\hline\hline
Case studied & Simulation results & MLJ formula \\ [0.5ex]
\hline
(1) 2-dim & 0.6276(3) & - \\
(2) $D_0$ & 0.7419(5)&0.7564 \\
(3) $D_1$ & 0.6022(2) & 0.6093 \\
(4) $D_2$ &0.6624(4) &  0.6716\\
(5) $D_3$ &0.6213(4) & 0.6283 \\ 
(6) $D_{\rm KP}$&0.6244(3)&0.6325\\[1ex]
\hline
\end{tabular}
\label{tab:lambda30}
\end{table}

With the above said, we can clearly see from table~\ref{tab:lambda30}
that the zeroth approximation of a constant diffusion coefficient
$D(x)=D_0$ gives far worse results for $D^*$ than all the proposed
expressions for coordinate-dependent $D(x)$. This observation supports
Zwanzig's idea that the modified FJ equation (\ref{eq:fickjacob2})
provides a better effective 1D description of the 2D diffusion problem
than the simple FJ equation (\ref{eq:fickjacob}). To further support
this conclusion, we plot in fig~\ref{fig:fig2} the function
$\Pi(x,t)=P(x,t)A_0/A(x)$ for $\lambda=30$ at $t=2\times 10^4$. The
plot shows the function $\Pi$ computed from simulations of the 2D
channel (blue circles), along with those computed from 1D FJ
simulations with $D_0$ (green stars), $D_1(x)=D_{\rm Z}(x)$ (Zwanzig's
formula - red diamonds), and $D_{\rm KP}(x)$ (Kalinay-Percus formula -
black pluses). The degree of agreement between the function $\Pi$ of
the 2D simulations and the approximations corresponding to $D_0$ (very
poor agreement), $D_1$ (significantly improved agreement), and $D_{\rm
  KP}$ (nearly perfect agreement), is clearly in accord with the
results for $D^*$ in table~\ref{tab:lambda30} showing precisely the
same trends. We do not show the PDF for the higher order
approximations ($D_2$, $D_3$) since, as evident from
fig.~\ref{fig:fig1}~(b), these expressions are derived from a
non-converging series expansion (see discussion in the previous
paragraph). In contrast, both $D_1=D_{\rm Z}$ and $D_{\rm KP}$ satisfy
$0<D(x)/D_0\leq 1$ [for any periodic function $A(x)$], which precludes
strong oscillations in $D(x)$ like the ones exhibited by $D_3(x)$ in
fig.~\ref{fig:fig1}~(b).  We thus conclude that although the 1D
projection method becomes less accurate for higher values of
$\epsilon$, there is still a significant improvement in the accuracy
of the PDF when $D_1(x)=D_{\rm Z}(x)$ and $D_{\rm KP}(x)$ are used
instead of the constant $D_0$.

\begin{figure}[tbh]
\centering\includegraphics[width=0.5\textwidth]{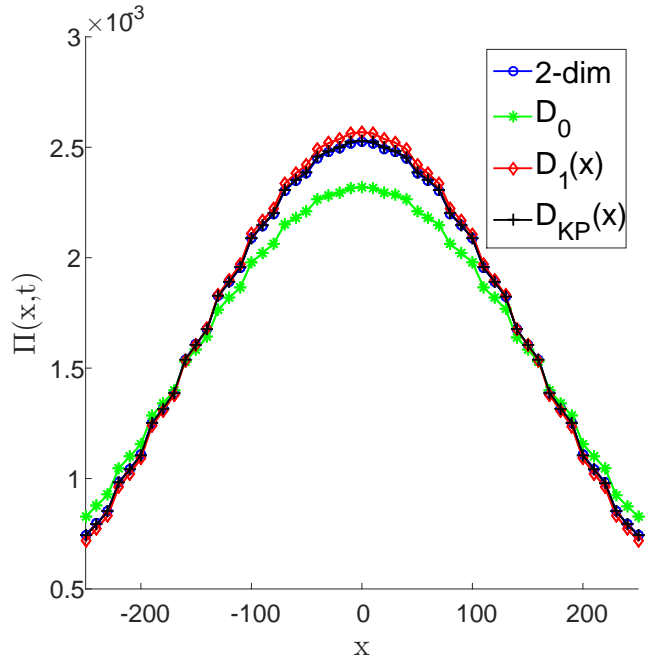} 
\caption{(Color Online) The numerical results for the
    function $\Pi(x,t)=P(x,t)A_0/A(x)$ computed from 2D channel
    simulation (blue circles), and from 1D FJ simulations with $D_0$
    (green stars), $D_1(x)$ (red diamonds), and $D_{\rm KP}(x)$ (black
    pluses) for $\lambda=30$. The data is collected at $t=2\times
    10^4$.}
  \label{fig:fig2}
\end{figure}

\section{Summary}
\label{sec:summary}

In this paper we revisited the problem of describing diffusive
dynamics along 2D periodic corrugated channels via a 1D FJ equation
with a spatially-dependent diffusion coefficient. In contrast to
previous attempts to derive expressions for $D(x)$ which were based on
steady state solutions, here we consider the non-stationary state of a
particle moving in an open channel. Similarly to the work in
ref.~\cite{kalinay06}, the expression derived here for $D(x)$ is in
the form of a series expansion in the parameter $\epsilon$ associated
with the aspect ratio of the channel; but in contrast to that work,
the formalism presented herein does not involve complicated
differnetal operators that are very hard to identify. The first order
approximation, $D_1(x)$, coincides with Zwanzig's formula for $D(x)$,
and for long wavelength channels (small $\epsilon$) it yields results
that are in perfect agreement with the 2D description. The agreement
is lost when $\epsilon$ is not sufficiently small, reflecting two
problems of the method. The first problem is a mathematical one: When
$\epsilon$ is not small, the series expansion does not converge
properly and cannot be truncated. The second problem is physical. The
1D description via FJ equation assumes a Markovian diffusion process,
which is only the case in the limit of fast relaxation of the
probability density in the transverese direction, i.e., for nearly
flat thin channels. Therefore, one should not be surprised by the
disagreement between the Langevin dynamics simulations of 1D periodic
systems and the 2D simulation results for short wavelength
channels. In fact, in this regime the 1D simulation results for the
effective diffusion coefficient do not even agree with the
Lifson-Jackson formula, which highlights yet another problem of the FJ
equation - its overdamped nature. FJ is a Smoluchowski equation which
is applicable only if, at length scale of the ballistic distance of
the dynamics, the variations in the force associated with the entropic
potential are much smaller than the characteristic friction force (see
discussion in \cite{sivan}) . This requirement is also not fulfilled
for short wavelength channels.

As a final note, we point out that the derivation of a series
expansion for $D(x)$ presented here for 2D channels can be extended to
three-dimensional (3D) geometries with cylindrical symmetry. In order
to do so, we assume that the 3D density, $\rho(x,r,t)$, has the same
form as the 2D density (\ref{eq:rho}) but with $y$ replaced by $r$,
and then substitute this form in the 3D diffusion equation written in
cylindrical coordinates. The resulting recurrence relation, which is
different than Eq.~(\ref{eq:rec1}) for the 2D problem, needs now to be
solved, and the steps of the derivation presented in section
\ref{sec:2dchannels} should be followed.


\newpage

\begin{thebibliography}{99}

\bibitem{hille} B. Hille, {\em Ion Channels of Excitable Membranes}\/
(Sinauer, Sunderland, Massachusetts, 2001).


\bibitem{oconnell} M. J. O'Connell (Ed.), {\em Carbon Nanotubes:
  Properties and Applications}\/, (CRC, Boca Raton, 2006).

\bibitem{schuring} A. Sch\"{u}ring, S. M. Auerbach, S. Fritzsche, and
R. Haberlandt, J. Chem. Phys. {\bf 116}, 10890 (2002).

\bibitem{weigl} B. H. Weigl and P. Yager, Science {\bf 283}, 346 (1999).

\bibitem{dekker} C. Dekker, Nat. Nanotech. {\bf 2}, 209 (2007).

\bibitem{jacobs} M. H. Jacobs, {\em Diffusion Processes}\/ (Springer,
  New York, 1967).

\bibitem{burada09} P. Sekhar Burada, G. Schmid, and P. H\"{a}nggi,
  Phil. Trans. R. Soc. A {\bf 367}, 3157 (2009).

\bibitem{bdb15} A. M. Berezhkovskii, L. Dagdug, and S. M. Bezrukov,
  J. Chem. O, J. Chem. Phys. {\bf 143}, 164102 (2015).

\bibitem{footnote3} The framework of Eq.~(\ref{eq:fickjacob2})
  employing a local diffusion coefficient is applicable for continuous
  functions $A(x)$ only, see discussion in M. Mangeat, T. Gu\'{e}rin
  and D. S. Dean, J. Chem. Phys. {\bf 149}, 124105 (2018).

\bibitem{zwanzig92} R. Zwanzig, J. Phys. Chem. {\bf 96}, 3926 (1992).

\bibitem{mangeat17} M. Mangeat, T. Gu\'{e}rin and D. S. Dean,
  J. Stat. Mech. Theory Exp. 123205 (2017).

\bibitem{reguera01} D. Reguera and J. M. Rubi, Phys. Rev. E {\bf 64},
  061106 (2001).

\bibitem{kalinay06} P. Kalinay and J. K. Percus, Phys. Rev. E {\bf
  74}, 041203 (2006).

\bibitem{kalinay05} P. Kalinay and J. K. Percus, J. Chem. Phys. {\bf
  122}, 204701 (2005).
  
\bibitem{kalinay14} P. Kalinay, J. Chem. Phys. {\bf 141}, 144101 (2014).

\bibitem{bradley09} R. M. Bradley, Phys. Rev. E {\bf 80}, 061142 (2009).
  
\bibitem{sivan} M. Sivan and O. Farago, Phys. Rev. E {\bf 98}, 052117
  (2018).

\bibitem{footnote1} Notice that in ref.~\cite{sivan}, we use a
  different notation involving the periodic function $\eta(x)$ and the
  constant $0\leq\epsilon<1$. These should be replaced here with
  $A(x)$ and $A_0$ via the relation $1+\epsilon\eta(x)=A(x)/A_0$.

\bibitem{lj} S. Lifson and J. L. Jackson, J. Chem. Phys. {\bf 36},
  2410 (1962).

\bibitem{footnote2} The scalig behavior $q_1\sim t^{-1/2}$ follows
  from $q_1\sim x/t$ and the fact that in diffusive dynamics $|x|\sim
  t^{1/2}$.

\bibitem{gjf} N. Gr{\o}nbech-Jensen and O. Farago, Mol. Phys. {\bf
  111}, 983 (2013).

\bibitem{gjf2} O. Farago and N. Gr{\o}nbech-Jensen, Phys. Rev. E {\bf
  89}, 013301 (2014).


  
\end{thebibliography}
\end{document}